# Intense sulphurization process can lead to superior heterojunction properties in Cu(In,Ga)(S,Se)$_2$ thin-film solar cells

*Oana Cojocaru-Mirédin[1,*], Elaheh Ghorbani[2], Mohit Raghuwanshi[1], Xiaowei Jin[3], Dipak Pandav[4], Jens Keutgen[1], Reinhard Schneider[3], Dagmar Gerthsen[3], Karsten Albe[2], Roland Scheer[5]*

[1]RWTH Aachen, I. Physikalisches Institut IA; Sommerfeldstraße 14, 52074 Aachen, Germany.

[2]*Fachgebiet Materialmodellierung, Institut für Materialwissenschaft, TU Darmstadt, Otto-Berndt-Straße 3, D-64287 Darmstadt, Germany*

[3]*Laboratory for Electron Microscopy (LEM), Karlsruhe Institute of Technology (KIT), Engesserstraße 7, 76131, Karlsruhe, Germany.*

[4]*Hochschule Anhalt Bernburger Straße 55, 06366 Köthen*

[5]Martin Luther-Universität Halle-Wittenberg, Institut für Physik, FG Photovoltaik, 06099 Halle.

**\*Corresponding author**: cojocaru-miredin@physik.rwth-aachen.de

**Abstract**

Sulphurization processes in Cu(In,Ga)Se$_2$ thin-film solar cells has been intensively studied in the last decade as a viable alternative to the existing Ga-grading. The main advantage of using S grading is that by substituting Se with S we will achieve not only an upshift of the conduction-band minimum as done by employing Ga grading, but also a downshift of the valence-band maximum. Several existing studies stipulate that S is very often inserted in too high concentrations into Cu(In,Ga)Se$_2$ absorber by sulphurization resulting in a deteriorated device performance instead of the expected beneficial effect. However, we demonstrate here that the intense sulphurization process when accompanied by Ga-grading leads to improved electrical properties of the buffer/absorber heterojunction. More exactly, this double grading at the absorber surface leads to strong reduction of the p-doping and hence to a change in the band diagram. This work also proves that the intense sulphurization process is accompanied by strong structural and chemical changes, i.e. by the formation of a S-rich CuIn(S,Se)$_2$ compound at the absorber surface. Finally, all these experimental findings were complemented by *ab-initio* calculations of the conduction-band and valence-band offsets between absorber and buffer obtained by using density functional theory. Hence, the present work opens up new possibilities for synthesizing Cu(In,Ga)(Se,S)$_2$ solar cells with superior cell performance when using an intense sulphurization process.

# 1. Introduction

Cu(In,Ga)(S,Se$_2$) (CIGSSe) is one of the most promising thin-film solar cell with a record conversion efficiency of 23.4%[1]. Yet, this value is still much lower than 33.7%, which is the theoretical value at the radiative limit[2]. There are several possible factors limiting the solar-cell performance, particularly nonradiative recombination via defect states taking place at the interfaces in the thin-film stack as well as at linear and planar defects. Therefore, in the last decades, researchers have put substantial efforts to considerably reduce nonradiative recombination processes in the solar cells.

One way to reduce the recombination activity and thus to increase the voltage without loss of the photocurrent is to introduce a distinct region with a larger band-gap energy than that of the absorber bulk in the top (towards buffer/absorber interface) and rear side (towards Mo/SLG interface, SLG stands from soda lime glass) of the absorber layer[3]. This has been achieved till now either by grading the relative Ga and In concentrations near the top and rear side of the CIGSe film[4] (so-called "double Ga-grading") or by introducing S in the absorber layer using a sulphurization process with increased S concentration at the top and rear side of the absorber layer[5-8]. The main advantage of using such a double grading (notch type) band-gap profile is that the top and rear sides of the absorber can be manipulated independently. More precisely, according to Ramanujam and Singh[9] an optimized front side grading improves the open circuit value ($V_{OC}$), whereas a back grading improves the product $V_{OC} \times J_{SC}$, where $J_{SC}$ represents the short-circuit current density. Optimizing the fill factor (FF), in addition, requires the accurate positioning of the grading notch[10]. In the present work we focus on the effect of the front-side grading on the buffer/absorber heterojunction properties. The novelty of this work is that we are investigating not only the effect of S grading with increased S concentration at the absorber surface, but also the co-effect of both, S and Ga, on the heterojunction properties.

Ga grading at the absorber surface leads to an increase in the conduction-band minimum (CBM)[11]. Yet, substituting S on Se sites leads not only to an upshift of the CBM[12], but also to a downshift of the valence-band maximum[11] (VBM lowered by 0.1 eV[13] owing to the reduced repulsion between Se-p and Cu-d states[14]). By lowering the VBM in the space charge region (SCR, close to the absorber surface), a hole-blocking barrier is generated and hence, interface recombination is reduced.[11, 15]. Apparently, this superior cell performance was still persistent when no or very little Ga was present in the S-rich region at the absorber surface[16].

Huang et al.[17] and Schubbert et al.[18] found that the presence of S at the surface of the absorber layer improves the cell performance (via higher open circuit, $V_{OC}$, value) if the S-rich region does not exceed the SCR width of approximatively 300 nm. They also argued that with greater depth and higher S concentration the performance of devices decreases (low fill factor, *FF*, and the reduction in current collection due to the formation of a notch in conduction band edge of the absorber). Yet, these authors did not provide any specific S concentration value for which the cell performance starts to degrade. To the best of our knowledge, such information is inexistent in the literature. Kobayashi et al.[19] suggested that inserting an S grading from ∼ 35 at.% to ∼ 10 at.% S within the first 30 nm from the CIGSe surface is enough to obtain superior cell performance. Yet, these values were measured using X-ray photoelectron spectroscopy, a technique which does not take into account the strong surface roughness of the CIGSe layer.

Hence, the goal of the present work is firstly to determine precisely the absolute S content in three dimensions at the absorber surface using the atom probe tomography (APT) capabilities. Moreover, the atomic structure of the S-rich region at the absorber surface is analyzed using high-resolution transmission electron microscopy (HRTEM). The dominant recombination site in the device is elucidated by $V_{oc}(t)$ transient measurements in combination with current-voltage analysis and doping density profiles from capacitance-voltage measurements. Finally, the band offset at the $In_xS_y$ buffer/absorber interface is theoretically calculated using density functional theory (DFT) calculations for four distinctive cases: $In_xS_y$:Na/CuIn$(S_{0.25}Se_{0.75})_2$, $In_xS_y$:Na/CuIn$(S_{0.50}Se_{0.50})_2$, $In_xS_y$:Na/CuIn$(S_{0.75}Se_{0.25})_2$, and $In_xS_y$:Na/CuInS$_2$ (i.e. a S content of 12.5 at.%, 25 at.%, 37.5 at.%, and 50 at.% at the absorber surface). Finally, a model of the role of sulfurization in the CIGSe absorbers is proposed based on the combination of experimental results and the DFT calculations.

## 2. Experimental procedure
### 2.1 Solar module fabrication
All samples were processed using the AVANCIS GmbH pilot line in Munich. The CIGSSe absorbers were grown with the SEL-RTP (SEL: stacked elemental layers; RTP: rapid thermal processing) using sulphurization of the absorber material[20]. With this process one can typically realize a double S-gradient at the top and rear side of the absorber layer. Moreover, a Ga gradient is also introduced at the rear side of the absorber typically leading to a Ga-free absorber surface. This is the reason why the absorber surface is named "CISSe" in the present paper. Before depositing the buffer layer, three different absorbers were synthesized as summarized in the Figure 1. For the first module (30x30 cm$^2$) type named "Standard CISSe" a moderate S content is present at the top and rear side of the absorber (Figure 1a), whereas for the second and third module type named "high CISSe" (Figure 1b) and "high CIGSSe" (Figure 1c) a much higher S content was introduced with the goal to study the effect of intense sulphurization on the module performance. For the third module type named "high CIGSSe", besides S, a higher Ga content was introduced at the top and the rear part of the absorber. We note that S and Ga are both present at the rear side of the absorber for all modules as schematized in Figure 1. $In_xS_y$:Na buffer layers (x and y can strongly vary upon the Na content in the buffer) were deposited onto the absorber with a dry physical vapor deposition process, using indium sulfide and a source containing a sodium compound. The cells within the module were completed by depositing the intrinsic ZnO and Al-doped ZnO layers. Cells were series-connected to modules via laser and mechanical patterning techniques.

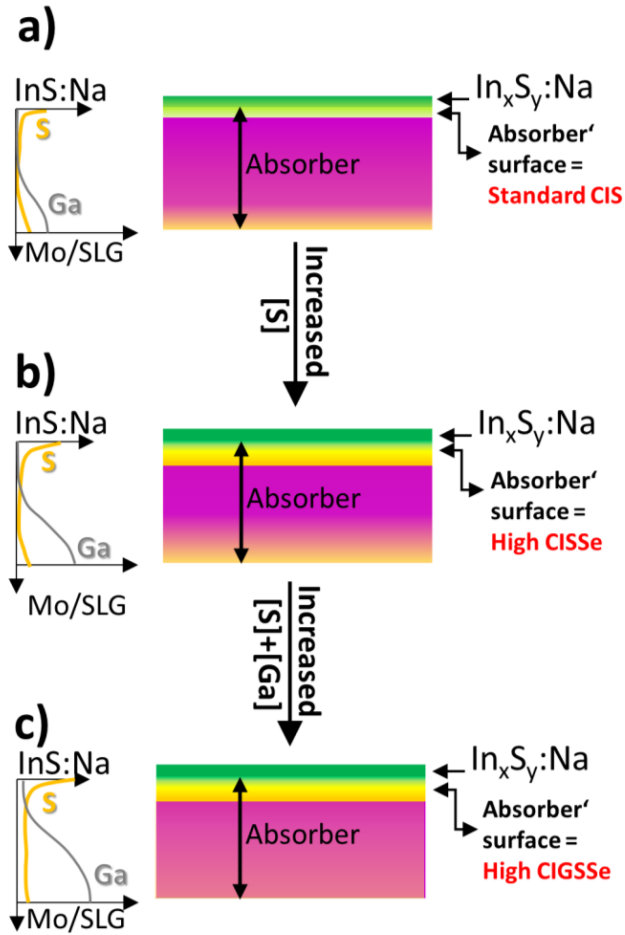

**Figure 1. Schematics of the cells fabricated in this work.** The schematics on the left side illustrates the S and Ga gradient in the three cells studied based on the glow discharge optical emission spectroscopy (GDOES, see Figure S1) measurements. On the right side, this schematic illustrates the S gradient in the three cell studied, especially at the absorber surface where the $In_xS_y$:Na buffer layer is present. The Standard CISSe cell (a) and the High CISSe cell (b) was synthesized with a double S-gradient at the top and rear side of the absorber layer. In the High CIGSSe cell (c) a much higher S content was inserted at the top part of the absorber, but at the rear side the S content is very weak. Based on GDOES experiments, the Ga content at the absorber surface for the Standard CISSe cell (a) and the High CISSe cell (b) is zero. Yet, for the high CIGSSe cell the Ga content was slightly increased at the absorber surface. Finally, all three cells contain a strong Ga gradient at the rear side of the absorber (illustrated only in the left-side, but not in the schematics on the right-side).

## 2.2 Analysis conditions

For the device characterization of 30x30 cm² sized laminated modules current-voltage characteristic (I-V) measurements are conducted with a Wacom XB-160T3W-5 constant light simulator under standard test conditions. The capacitance-voltage measurements were performed using Agilent E4980A LCR meter under dark and at room temperature.

The time-dependent $V_{oc}(t)$ measurements were performed at 298 K on non-laminated devices using active temperature control. Initially, the samples were relaxed in darkness for 16 hours at a temperature of 318 K in the light-tight chamber. Red light illumination was realized by a 590 nm long pass filter in the light path of the Xenon type sun simulator. Opening the light shutter starts the measurement of $V_{oc}(t)$ by a Keithley 2400 digital source meter in 2 seconds intervals. The sample temperature was measured on a witness sample next to the specimen. We performed a voltage-temperature correction (using additional $V_{oc}(T)$ data) after the measurement to account for the error introduced by temperature fluctuations at the start of the experiment. The $\Delta V_{oc}(t)$ was referenced to the $V_{oc}$ value 100 s after opening the light shutter. More details of $V_{oc}(t)$ measurement can be found in Ref.[21] Also current-voltage measurements were carried out in this setup with (red light) and without (white light) filter simulating 0.5×AM1.5 and 1.0×AM1.5 conditions, respectively.

For the transmission electron microscopy (TEM) investigations, cross-sectional TEM specimens were prepared by focused-ion-beam (FIB) milling with a Thermo Fisher dual-beam Helios G4 FX microscope. TEM lamellae were prepared by FIB milling with a $Ga^+$-ion beam at 30 kV until the sample thickness reached about 200 nm. The FIB lamellae were then attached to Si lift-out grids. Ion-beam milling at a voltage of 5 kV and an ion current of 12 pA was applied to further reduce the lamella thickness down to approximatively 100 nm. Finally, an ion-beam polishing with a low voltage of 1 kV and a low current of 7 pA was carried out to minimize ion-induced amorphization and Ga implantation.

High-resolution TEM (HRTEM) studies were carried out using a FEI Titan[3] 80-300 transmission electron microscope operated at 300 kV. The instrument is equipped with a thermally assisted field-emission electron gun and an aberration corrector in the imaging lens system. In the present work, the microstructural characterization of the solar cells was performed by diffraction-contrast TEM imaging, whereas the crystal structure was elucidated at atomic resolution by HRTEM. TEM images were recorded by means of 4 mega-pixel CCD camera (Gatan UltraScan 1000 P) with typical exposure times of a few seconds. Electron diffraction patterns were calculated using the JEMS software[22] for comparison with the experimental data obtained by two-dimensional Fourier transformation of HRTEM images.

APT analysis was performed using a local electrode atom probe (LEAP 4000 Si, CAMECA instruments). The specimen was maintained at 50 K temperature under ultra-high vacuum conditions ($10^{-11}$ mbar). Field evaporation was triggered using green laser (532 nm wavelength) pulses of energy 3-5 pJ with 200-250 kHz repetition rate. The detection rate of 0.5% (on an average detection of 1 ion every 200 pulses) was kept constant throughout the analysis.

Calculations in the framework of density functional theory (DFT) were performed using the VASP package[23] with projector augmented wave pseudopotentials[24]. A planewave cut-off energy of 400 eV was used for all calculations. Γ-centered *k*-point meshes of 4 × 4 × 2 and 2 × 1 × 8 were used for 16-atom chalcopyrite unit cells and 36-atom $NaIn_3S_5$ orthorhombic unit cell, respectively. The exchange-correlation potential of choice is the Heyd-Scuseria-Ernzerhof (HSE)[25] with 25% of exact exchange added to 75% semi-local exchange of Perdew-Burke-Ernzerhof (PBE)[26]. The screening length was set to 0.13 Å$^{-1}$, which gives an excellent estimation of band gap for chalcopyrite systems[27]. For alloying $CuInSe_2$ with S (Ga), S and Se (In and Ga) were distributed uniformly along the [001] direction of the mixed anion (cation) alloys. Our calculated band gaps are 1.10, 1.58, 1.41, 1.69, and 2.47 eV for $CuInSe_2$, $CuIn(S_{0.5}Se_{0.5})_2$, $Cu(In_{0.5}Ga_{0.5})Se_2$, $Cu(In_{0.5}Ga_{0.5})(S_{0.5}Se_{0.5})_2$ and $NaIn_3S_5$, respectively.

To determine the band offset between materials X and Y ($\Delta E_v(X/Y)$), the band edges were aligned with respect to the core level energies according to the equation: $\Delta E_v(X/Y) = \Delta E_{v,c'}(Y) - \Delta E_{v,c}(X) + \Delta E_{c,c'}(X/Y)$, where $\Delta E_{v,c}(X)$ and $\Delta E_{v,c'}(Y)$ are the energy differences between the VBM and core level energy of the X and Y compounds, respectively. To the best of our knowledge, the interfacial orientation relation between studied compounds is not available. Therefore, our calculations focus on the natural band line-ups between the absorber (such as $CuInSe_2$, $CuIn(S_{0.5}Se_{0.5})_2$ abbreviated $CuIn(S,Se)_2$, $Cu(In_{0.5}Ga_{0.5})Se_2$ abbreviated $Cu(In,Ga)Se_2$, and $Cu(In_{0.5}Ga_{0.5})(S_{0.5}Se_{0.5})_2$ abbreviated $Cu(In,Ga)(S,Se)_2$) and the $NaIn_3S_5$ buffer. Therefore, the term $\Delta E_{c,c'}(X/Y)$ in the equation above was considered to be zero. The conduction band offsets were then calculated by using the relation $\Delta E_c(X/Y) = \Delta E_g(X/Y) - \Delta E_v(X/Y)$, where $E_g(Y)$ and $E_g(X)$ denote band gaps of Y and X structures, respectively. The Cu *2s* energy was used as reference energy levels to align band edges of $CuIn(S,Se)_2$, $Cu(In,Ga)Se_2$ and $Cu(In,Ga)(S,Se)_2$ with respect to $CuInSe_2$.

The band alignment between $CuInSe_2$ and $NaIn_3S_5$ was performed by applying the In *2s* energy as reference core energy in the equation above. Note that energy band alignment between $CuIn(S,Se)_2$, $Cu(In,Ga)Se_2$, $Cu(In,Ga)(S,Se)_2$ absorbers and $NaIn_3S_5$ buffer can be derived from the equation above by making use of the transitivity rule, $\Delta E_v(Y/Z) = \Delta E_v(X/Z) - \Delta E_v(X/Y)$, where X represents $CuInSe_2$ and Y and Z represent $CuIn(S,Se)_2$, $Cu(In,Ga)Se_2$, $Cu(In,Ga)(S,Se)_2$ or $NaIn_3S_5$.

## 3. Results and Discussion

### 3.1 Module (Cell) performance

Intense sulfurization of the absorber layer degrades the module and, hence, cell performance exhibiting a reduction in conversion efficiency. Yet, by subsequently increasing the Ga content at the absorber surface the conversion efficiency increases drastically exceeding the record efficiency of standard CISSe cell. More specifically, Figure 2 shows that by increasing the S content in the absorber the $V_{OC}$, *FF*, and the short circuit current ($J_{SC}$) decrease, while the series resistance ($R_S$) increases. However, by subsequently increasing the Ga content at the absorber surface, except for the $J_{SC}$, all the other cell parameters are restored and even improved when comparing them with the standard CISSe cell.

According to the equation $V_{OC} = \frac{AkT}{e}\ln(\frac{J_{ph}}{J_0} + 1)$, where A is the diode quality factor, the open circuit voltage $V_{OC}$ is a measure of the recombination occurring in the solar cell[28] and the effective absorber bandgap (both reflected in the saturation current density $J_0$). Figure 2a shows that the $V_{OC}$ value decreases with intense sulfurization, but increases drastically when Ga is introduced at the absorber surface. Contrary to $V_{OC}$, the $J_{SC}$ parameters degrade when adding S and Ga (cf. Figure 2c). This can be explained by two reasons. First, by increasing the amount of S and/or Ga at the absorber surface the minimum band gap in the absorber increases (so-called 'band grading'[29]) reducing the photon absorption and thus $J_{SC}$. Second, according to Ramanujam and Singht[9], a reduced $J_{SC}$ value can be due to the non-optimized S and Ga content at the rear side of the absorber, which leads to deteriorated absorber/Mo-interface properties. For example, Insignares-Cuello et al.[6] observed the formation of a $Mo(S,Se)_2$ alloy at the absorber/Mo interface during the sulphurization process, which is accompanied by a reduced $J_{SC}$ value (from 35.8 to 33.2 mA/cm$^2$). Finally, the variation in FF between these three samples appears to be connected with the series resistance $R_S$, i.e. the increase in series resistance $R_S$ is accompanied by the reduction in FF observed after increasing the S content in the absorber layer as shown in Figure 2b. Yet, by increasing the Ga content at the surface of the absorber the series resistance $R_S$ is drastically reduced, restoring and even improving the FF value of the cell.

All in all, the overall module performance is degraded when increasing the S content in the absorber. However, this module performance is restored and even improved when Ga is subsequently added to the absorber surface. Figure 2c shows that the efficiency of the CISSe cell is indeed reduced by 6 % (normalized value) when S is found in excess in the absorber layer, and the efficiency is interestingly improved by 1 % (normalized value, small but realistic change since efficiency measured on big module containing many cells) when Ga is subsequently added at the CISSe surface. Hence, understanding the impact of S and Ga on the efficiency becomes crucial for the improvement of solar cell efficiency.

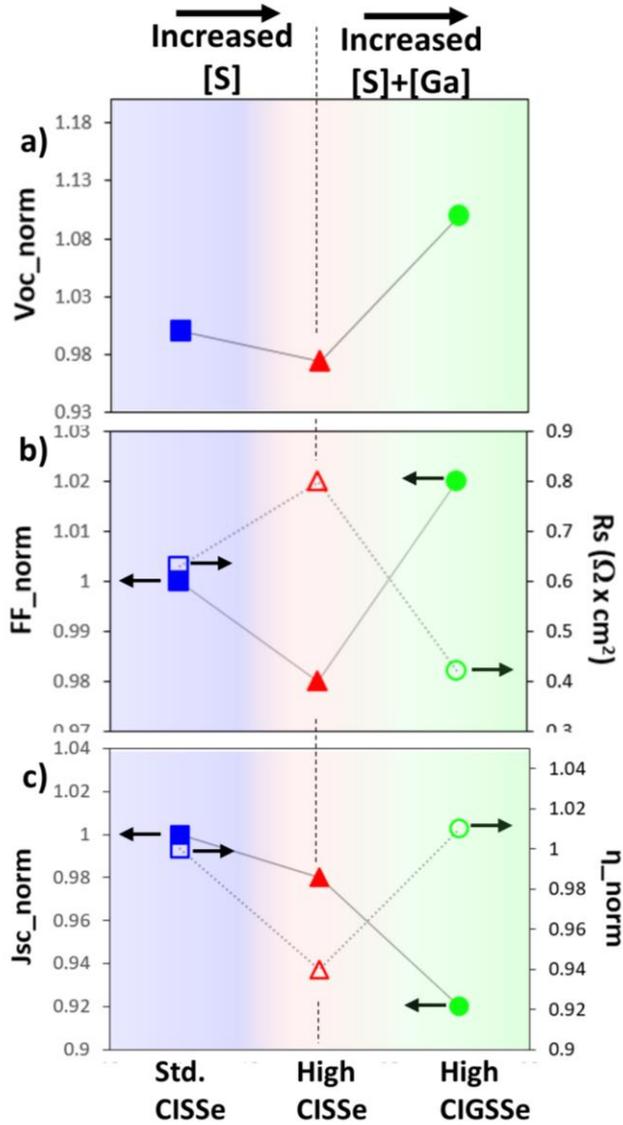

*Figure 2. Solar parameters for the investigated modules.* *These cell parameters (i.e. open circuit voltage ($V_{OC}$), fill factor (FF), short circuit current ($J_{SC}$), and series resistance (Rs)) degrade by increasing the sulphur (S) content at the $CuIn(S,Se)_2$ (CISSe) surface, while, except for the $J_{SC}$ they get restored and even improved by subsequently increasing the gallium (Ga) content at the absorber surface. This leads to a higher conversion efficiency η measured for the high CIGSSe than that measured for standard CISSe and high CISSe. Note that all lines are guides to the eye to highlight the trends. Some of the values given in this figure are normalized to the values of the standard CISSe.*

## 3.2 Net doping and recombination

The profiles from Figure 3 show the net acceptor concentration $N_A$ (or p-type doping) versus the depletion width $d(U) = \sqrt{\frac{2\varepsilon_0\varepsilon_r}{q} \cdot \frac{1}{N_A} \cdot (V_{bi} - U)}$ calculated using the depletion approximation model. The net acceptor concentration $N_A$ decreases when the S or S + Ga content increases in the CISSe absorber surface. Moreover, the width of the SCR, $d(U)$, had doubled with increasing the S and Ga content, which agrees well with the calculated acceptor concentration $N_A$, since lower doped material will have a wider depletion region and vice versa.

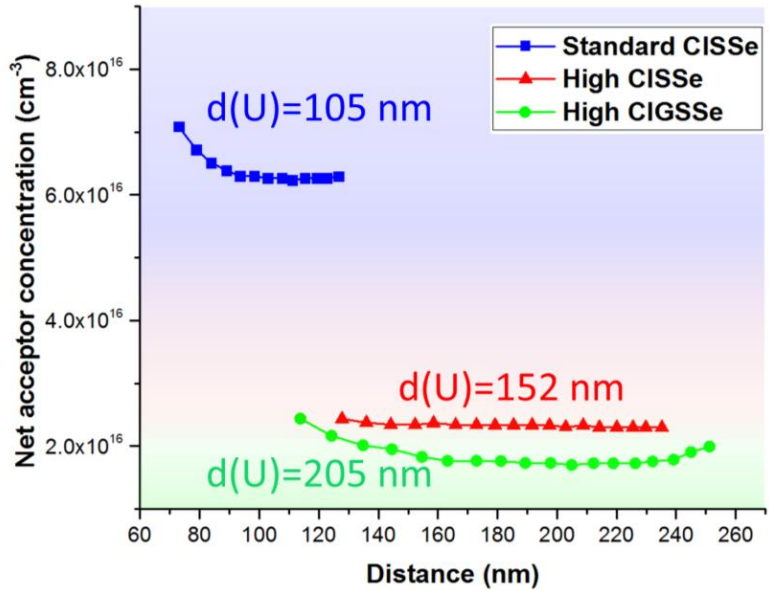

*Figure 3. Calculated net acceptor concentration for the three investigated modules. These profile obtained from the C-V profiles shows a big difference in net acceptor concentration between the standard CISSe module and high CIGSSe modules. The obtained net acceptor concentration for the intermediate sample, high CISSe modules, is very close to the calculated value for high CIGSSe. We note here that the net acceptor concentration values were calculated considering a one-sided heterojunction. Based on the shown net acceptor concentration, the depletion width d was calculated using the depletion approximation: $d(U) = \sqrt{\frac{2\varepsilon_0\varepsilon_r}{q} \cdot \frac{1}{N_A} \cdot (V_{bi} - U)}$, where $V_{bi}$ is the bias voltage.*

In order to test if interface or rather bulk recombination is dominating the current transport, $V_{oc}(t)$ transients were recorded under red light illumination (see Ref.[30]). For these experiments the well-known red-light induced doping increase is exploited. Figure 4a gives the transients for the standard CISSe and high CIGSSe samples. The falling $\Delta V_{OC}$ of the standard CISSe sample under red light illumination (dashed line) indicates that this device is limited by *interface recombination*: the interface recombination rate increases with illumination time due to the decrease of $E_{p,z=0}$ (see Figure 9) upon illumination induced doping increase. Interface recombination under red light illumination is also indicated by the kink in the JV curve of this sample (marked by the arrow in Figure 4b). However, application of white light (full line) both strongly changes the $\Delta V_{oc}(t)$ towards stable values and removes the JV kink. The reason could be the photo induced doping of the $In_xS_y$ buffer layer, similar to what has been reported for CdS buffer layers[31]. Such buffer doping increases $E_{p,z=0}$ but is missing by exclusive red light illumination.

In contrast to the standard CISSe sample, the high CIGSSe sample exhibits a rising $\Delta V_{oc}(t)$ curve and does not show a kink under red light JV measurement (green dashed line). This is a strong indication for *bulk recombination* limiting this device with the two possibilities of space-charge-region and quasi-neutral region recombination. The same is found for the high CISSe sample as depicted in the plots of Figure S2.

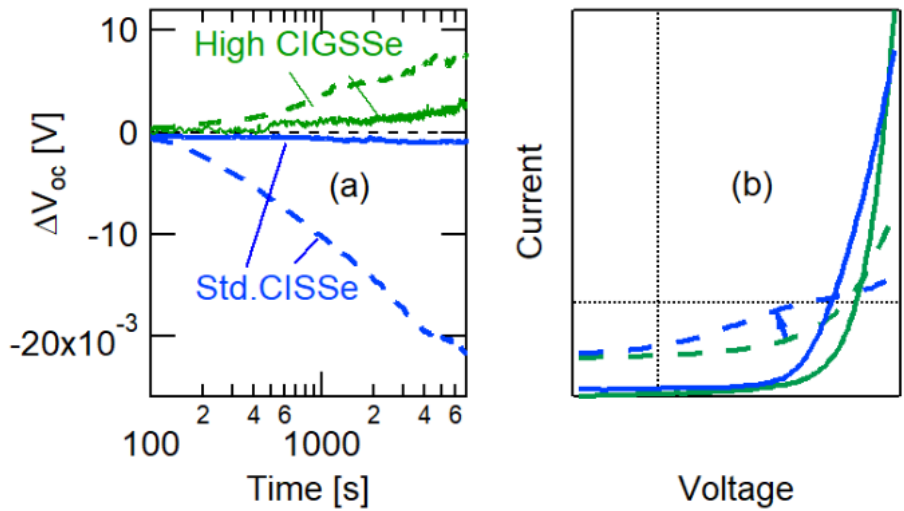

*Figure 4. $V_{oc}(t)$ transient measurements of the investigated cells.* *(a) Open circuit voltage transients of the Standard CISSe and high CIGSSe samples given as $\Delta V_{oc}(t)$ under red light illumination (dashed lines) and white light illumination (full lines). (b) Current-Voltage curves under red light (dashed lines) and white light (full lines) illumination. The arrow indicates the position of the kink in the red light JV curve of the standard CISSe cell.*

Hence it may be postulated that even under white light illumination, the standard CISSe device has some larger contribution of interface recombination compared to both other samples with higher S at the surface. The combined C-V and $V_{OC}(t)$ investigations could point towards an electronic reason behind the improved $V_{OC}$ value when increasing both the S and Ga content at the CIGSe surface. However, in addition there may be differences in structure and/or composition. Hence, in the following subsection 3.3, we first give a detailed description of the atomic structure and composition of the heterointerface. Then, these experimental results are combined with the theoretical results obtained by DFT calculations (described in 3.4) in order to propose a model which can explain the effects of S and Ga addition into the surface of the CISe layers on the device performance of the corresponding solar cells.

### 3.3 Structure and composition of the buffer/absorber interface at the nanoscale

Figure S3 reveals the structural details of the corresponding absorber/buffer/ZnO interfacial regions in all three solar cells. The $In_xS_y$:Na buffer layer has the thickness of approximately 50 nm. The absence of strong local image-contrast variations in the TEM bright-field images suggests that the buffer has either a nanocrystalline or amorphous structure.

To investigate more precisely the structure of the $In_xS_y$:Na buffer, HRTEM investigations at the $In_xS_y$:Na buffer/absorber interface were conducted (cf. Figure 5). The $In_xS_y$:Na buffer layers of all samples exhibit a nanocrystalline structure with grain sizes of a few nanometers. The $In_xS_y$:Na nanocrystallites are randomly oriented and, thus, do not reveal any orientation relationship with the adjacent CISSe grains. In the Fourier-transformed HRTEM images of the buffer layers (upper right corner in Figure 5a,b), a large number of reflections with different spatial frequencies $g$ can be seen. Exemplarily, the spatial frequencies of two reflections $g_1 = 1.60$ nm$^{-1}$ and $g_2 = 3.06$ nm$^{-1}$ are marked and compared with values calculated by means of the JEMS software[22]. For comparison with the experimentally determined spatial frequencies, three different phases were considered that could be present in the buffer layers. These phases are $In_2S_3$ with the tetragonal structure ($I4_1/amd$, lattice parameters a = b = 0.76 nm, c = 3.24 nm[32]), $In_2S_3$ with the cubic structure (Fd-3m, lattice parameters a = 1.08 nm[33]) and $NaIn_3S_5$ with the orthorhombic structure (Pbam, lattice parameters a = 1.25 nm, b = 1.61 nm, c = 0.38 nm[34]). As a result, the spatial frequency $g_1$ is consistent with the (111) planes of $In_2S_3$ (1.61 nm$^{-1}$) with cubic structure, the (103) planes of $In_2S_3$ (1.61 nm$^{-1}$) with tetragonal structure and the (200) planes of $NaIn_3S_5$ (1.60 nm$^{-1}$). The spatial frequency $g_2$ corresponds to the (113) planes of cubic $In_2S_3$ (3.08 nm$^{-1}$), the (213) planes of tetragonal $In_2S_3$ (3.09 nm$^{-1}$) or the (201) planes of $NaIn_3S_5$ (3.08 nm$^{-1}$). In addition to $g_1$ and $g_2$, reflections with 5.25 nm$^{-1}$ were detected corresponding to the (044) planes of the cubic $In_2S_3$, the (2,2,12) planes of tetragonal $In_2S_3$, or the (002) planes of $NaIn_3S_5$ (5.26 nm$^{-1}$). We note that the most intense (110) reflection of $NaIn_2S_5$ at about 1.01 nm$^{-1}$ is absent in the FT HRTEM images. However, this is not surprising because FT HRTEM images were evaluated and not electron diffraction patterns. Reflections at small spatial around 1.01 nm$^{-1}$ will, if at all, only show up with weak intensities in FT HRTEM images due the bad contrast transfer of the objective lens at small spatial frequencies at small defocus values that are chosen for HRTEM imaging.

Hence, all three considered phases could be present in the buffer layer. However, it has been recently reported that tetragonal $In_2S_3$ is much more stable than cubic $In_2S_3$ below 476°C[35]. As the deposition temperature of our $In_xS_y$:Na buffer layer was much lower than 476°C, the presence of tetragonal $In_2S_3$ within the buffer is more likely than that of cubic $In_2S_3$. In addition, the phases of wurtzite CuS (P6$_3$mc, lattice parameters a = b = 0.38 nm, c = 1.63 nm)[36] and wurtzite CuSe (P6$_3$/mmc, lattice parameters a = b = 0.40 nm, c = 1.73 nm)[37] are also compatible with the reflections in Figure 5a. It should be noted that the presence of other phases cannot be completely excluded due to the small analyzed volume.

Figure 5c shows a HRTEM image of the interfacial region between the $In_xS_y$:Na buffer layer and the absorber for the high CIGSSe cell. The Fourier transforms of image regions (i) and (ii) are inserted in the lower-left and upper-right corners. Region (i) is located directly at the absorber surface whereas region (ii) starts at a distance of 5 to 10 nm below the buffer and extends further into the absorber layer. We note that chemical analyses by APT measurements presented in the following yield a Ga content of only 1.7 at.%. Hence, Ga-containing compounds were not considered for the phase assignments based on the evaluation of the Fourier-transform patterns. The Fourier transform of image region (i) was found to be compatible with chalcopyrite $CuInS_2$ (I-42d, lattice parameters a = b = 0.55 nm, c = 1.11 nm)[38]. From the accuracy of the lattice-parameter measurements, a Se content of up to 25 at.% percent ($CuIn(S_{0.5}Se_{0.5})_2$ with lattice parameters a = b = 0.56 nm, c = 1.14 nm)[39] cannot be completely excluded. The analysis of the Fourier-transformed image region (ii) yields lattice parameters, which fit well with $CuInS_2$ or S-rich $CuInSe_2$ recorded along the [110] zone axis. Thus, based on the HRTEM investigations $CuInS_2$ or S-rich $CuIn(S,Se)_2$ at the absorber surface was detected for the high CIGSSe sample. For the other two cells, namely standard CISSe and high CISSe, the Fourier-transform patterns obtained at the absorber surface only fits to $CuInSe_2$, indicating pure $CuInSe_2$ or S-poor $CuIn(S,Se)_2$ phase.

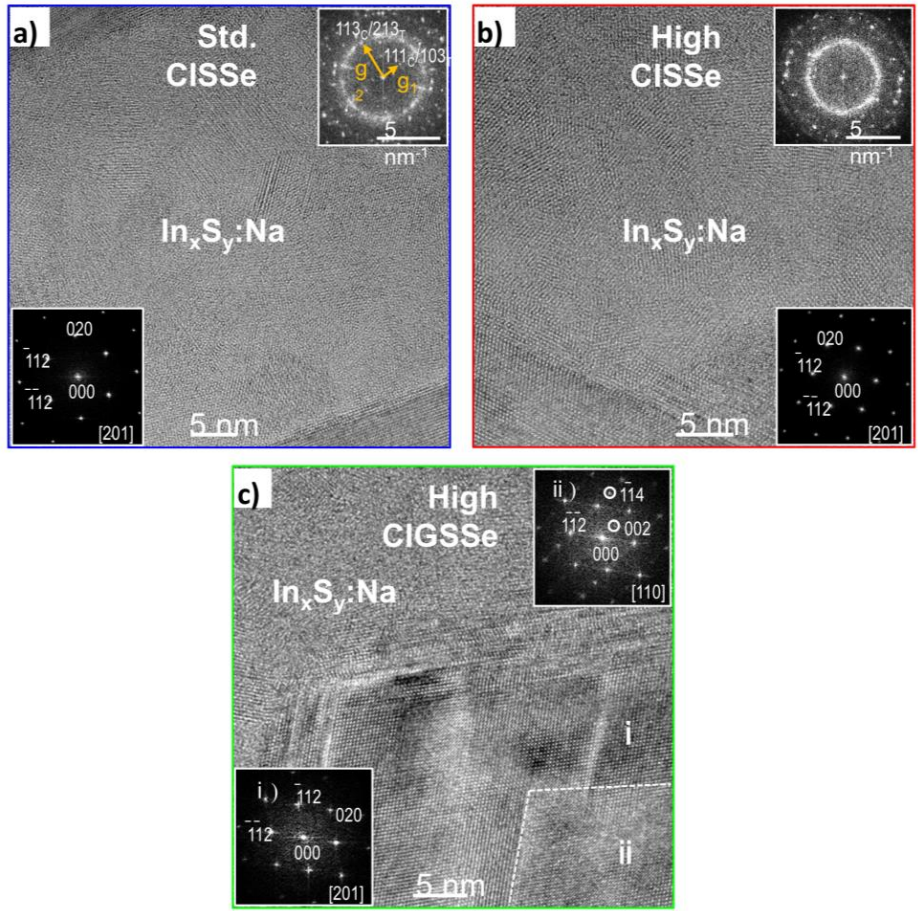

**Figure 5. Cross-section HRTEM images of the buffer/absorber-interface region.** The interfacial region between the $In_xS_y$:Na buffer and absorber is shown for a) the standard CISSe, b) the high CISSe and c) the high CIGSSe cell. The images were acquired along the [201] zone axis of the absorber as revealed by the Fourier-transform patterns shown either in the lower left corner for a) or in the lower right corner for b). The inset in the upper-right corner of a) and b) shows the Fourier-transformed image section of the $In_xS_y$:Na buffer layer, where the selected spatial frequencies $g_1$ and $g_2$ are marked and discussed in the text. The insets in the lower-left and upper-right corners in c) show the Fourier-transformed regions i) and ii) of the absorber.

APT studies were conducted on these three cells to determine precisely the composition of the $In_xS_y$:Na/CIGSe heterointerface and to compare with the standard CdS/CIGSe heterointerface[40-42]. Moreover, these APT studies will clearly demonstrate if $CuInS_2$ is present at the CIGSe surface of the high CIGSSe cell and will determine its exact composition. Finally, by combining these APT results with the previously presented C-V and $V_{OC}(t)$ results we can clearly reveal how the cell performance changes based on the S content incorporated at the surface of the absorber through sulphurization.

Figure 6 summarizes one representative APT investigation obtained for each sample. Surprisingly, the interface between the buffer and the absorber is very rough with an average roughness value ranging from 60 to 200 nm. This accentuated interface roughness observed in this work is unusual and it had never been observed in our previous APT studies[40-41, 43-45] performed on non-sulphurized CIGSe layers. Hence, it is believed that this unusually-high roughness observed here comes from the sulphurization process in agreement with Kim et al.[11].

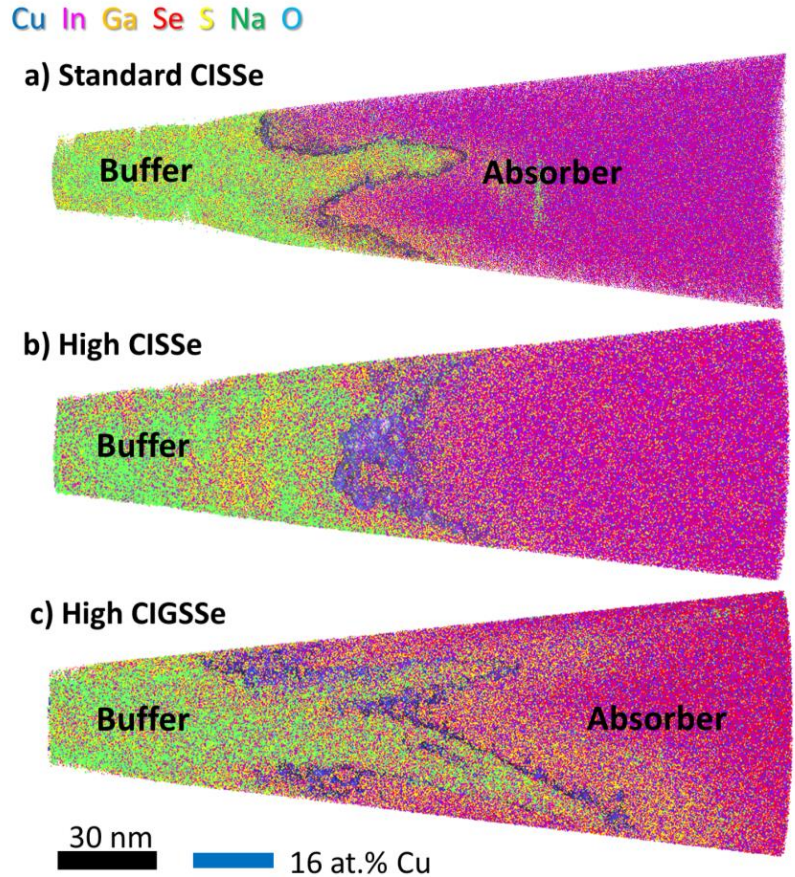

**Figure 6. APT investigation of the $In_xS_y$:Na/CIGSe interface.** 3D map showing the elemental (Cu, In, Ga, S and Se) and impurity (Na and O) distribution at the $In_xS_y$:Na/CIGSe interface for a) standard CISSe, b) high CISSe, and c) high CIGSSe cells. The interface roughness is very pronounced for all three samples. A S-rich region (pronounced yellow color in Figure 6c), detected 30 nm deep inside the absorber layer, is clearly visible for high CIGSSe cells. Yet, S was also observed at the absorber surface for the other two cells, standard CISSe and high CISSe, but not in such high magnitude and limited in a very narrow region.

Surprisingly, the $In_xS_y$:Na buffer composition differs between the three investigated samples. For the standard CISSe cell the buffer composition approaches the composition of the $In_2S_3$ (more exactly the cationic and anionic composition equals to ~40 at.% and ~60 at.%, respectively; see Table 1), whereas for the other two samples the buffer composition deviates from $In_2S_3$ (the cationic and anionic composition approaches ~50 at.% each). Yet, from the TEM results we know that the $In_2S_3$ structure is preserved for the buffer layer of all three investigated cells. This discrepancy can be explained by *i)* surplus of Na is inserted as interstitials in $In_2S_3$ and *ii)* the possible interdiffusion between the buffer and the various S-containing absorbers during the deposition of subsequent layers such as ZnO. The latter is confirmed by the increase in Se content and the decrease in S content in the buffer layer when going from moderate to high S content, which is contrary with what have been observed for the cells with S-free absorbers[41].

*Table 1. Buffer composition in all three cells obtained by APT. The buffer composition averaged on three different APT measurements shows that the S content in the buffer layer decreases from the standard sample to the high CIGSSe sample. The composition given in this table is accurate, i.e. the peak decomposition algorithm[46] was applied for the peaks which were overlapping on the mass spectrum. The statistical errors are not given here since they are very small (do not exceed 0.1 at.%).*

|         | **Element** | **Std. CISSe** | **High CISSe** | **High CIGSSe** |
|---------|-------------|----------------|----------------|-----------------|
| **Cations** | In (at.%)   | 32.5 | 34.5 | 34.1 |
|         | Cu (at.%)   | 1.7  | 1.2  | 1.4  |
|         | Na (at.%)   | 5    | 10.8 | 10.2 |
| **Anions**  | S (at.%)    | 58.6 | 51   | 51   |
|         | Se (at.%)   | 2.2  | 2.5  | 3.3  |

From the 3D maps shown in Figure 6, one can argue that a non-negligible amount of S is detected at the surface of the absorber for all three cells. Yet, for the high CIGSSe cell this S-rich region is surprisingly extended to approximatively 30 nm deep inside the absorber layer, which might indicate the presence of a S-rich phase in agreement with the HRTEM results from Figure 5c. To find out the exact composition of S at the CIGSe surface, proximity histograms (or proxigrams)[47] were constructed, which represent the most advantageous determination of S composition in a region delimited by a very rough interface.

Figure 7 shows these proxigrams constructed for the $In_xS_y$:Na/absorber interface of all three cells. Yet, the peak overlap observed for example between $^{160}S_5^+$ and $^{160}Se_2^+$, $^{80}S_5^{2+}$ and $^{80}Se^+$, and several isotopes of $Se_2^{2+}$ with $Se^+$ do not allow to obtain accurate composition profiles using standard software IVAS. Therefore, a new software called EPOSA$^{TM}$ was specially designed to automatically correct the composition on the proxigram using the peak-decomposition algorithm[46]. Hence, all composition profiles given in Figure 7 are obtained using the EPOSA[46] software and are, hence, accurately corrected.

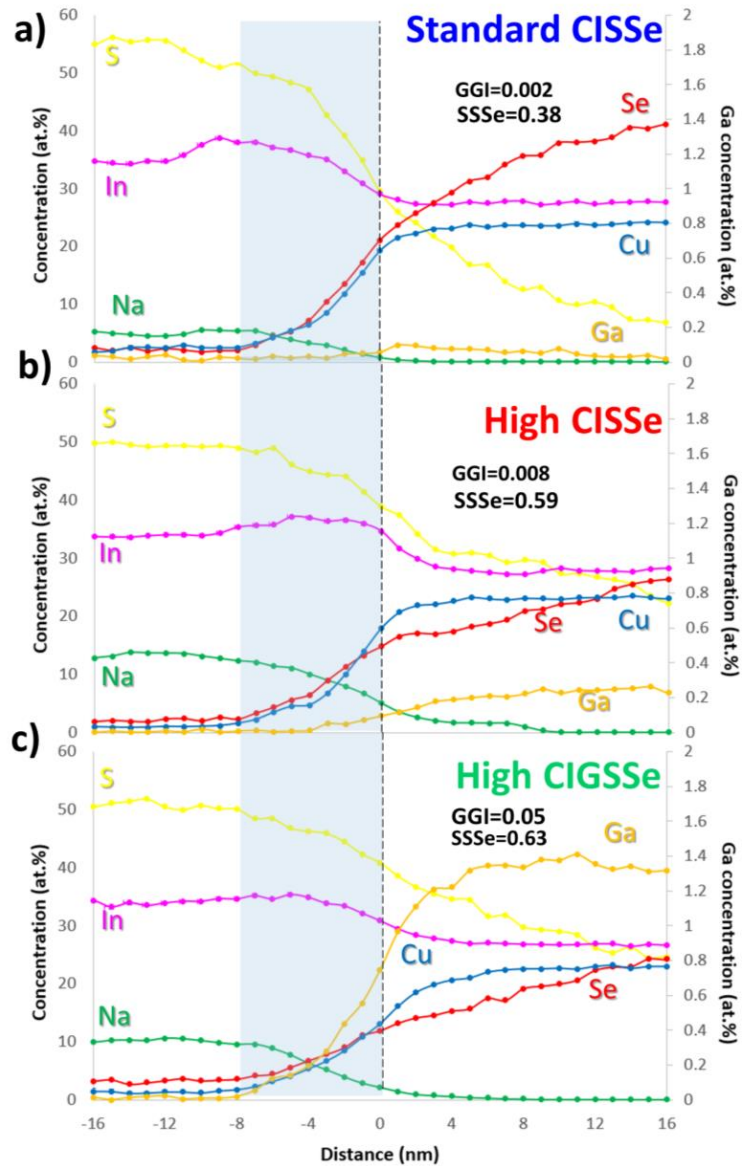

**Figure 7. Composition of the buffer/absorber interface.** The S (yellow), In (pink), Na (green), Se (red), Cu (blue), and Ga (mustard yellow) linear compositional profiles are constructed at the buffer/absorber interface using the proxigram procedure for a) standard CISSe, b) high CISSe, and c) high CIGSSe cells. Note, that the Ga-concentration axis is given at the right-hand side. The blue-shaded region indicates the region where strong Cu and Se diffusion was observed close to the hetero-interface. To build these proxigrams, a Cu iso-concentration surface was constructed using a Cu iso-concentration value of 16 at.%.

These proxigrams prove that the S gradient at the absorber surface is very steep for the standard CISSe cell reaching a very low concentration already after 15 nm, whereas for the other two samples the S content at the absorber surface decreases only gradually expanding deeply into the absorber layer (not shown here). Moreover, the number of S atoms increases with the decreasing the number of Se atoms, suggesting that the S atoms are substituted by the Se atoms at the surface of the absorber layer, in agreement with Kobayashi et al.[19].

The averaged S and Ga composition at the surface of the absorber is given for all three samples in Table 2. It can be clearly seen that the S content at the surface increases considerably from one sample to another, i.e. from ~17 at.% for standard CISSe cell to ~ 34 at.% for high CIGSSe resulting in an enormous difference in S composition of 17 at.% at the absorber surface; whereas the S is almost absent in the absorber close to the back-interface (see Figure S1b). The Ga content, however, does only vary very little between cells with a difference in Ga composition of maximum ~1.6 at.% at the absorber surface (see Table 2). At the back interface, the difference in Ga content between Standard CISSe (~20 at.% Ga; measured by APT) and High CIGSSe (~22,5 at.% Ga; measured by APT) is only slightly higher than that registered at the surface; i.e. of ~2.5 at.%, but still far below the 17 at.% deviation in S composition registered at the absorber surface. Therefore, we conclude here that the major impact on the cell performance is played by S and not by Ga at the absorber surface; although the Ga grading plays naturally a vital role for a good performing cell.

In agreement with the TEM investigations, the S content for the high CIGSSe cell is surprisingly higher than the S content in high CISSe cell. Most probably, Ga favors S diffusion in the absorber material. One needs to mention here that the APT results reveal only small Ga contents for the high CIGSSe cell with a GGI value of approximatively 0.06. Yet, the Ga content measured in the other two samples, standard CISSe and high CISSe, is almost 8 times smaller with GGI values lower than 0.008 (see Table 2). Moreover, such high S content detected at the surface of the absorber for the high CIGSSe cell confirms the observation made using HRTEM: formation of the $CuInS_2$ phase at the absorber surface. APT results even complement the HRTEM results by proving that this $CuInS_2$ compound contains ~15 at.% Se. The formation of $CuInS_2$ phase at the absorber surface is in agreement with the recent APT work of Aboulfadl et al.[48].

*Table 2: Ga, In, Se, and S content at the absorber surface for all three cells. The Ga, In, S, and Se contents at the absorber surface (first 15 nm) and the corresponding GGI and SSSe ratios (GGI=[Ga]/([Ga]+[In]) and SSSe=[S]/([S]+[Se])) are given for three independent APT measurements for all three samples. The statistical errors are not given here since they are very small (below 0.1 at.%).*

| Sample | Measurement nb. | [Ga] (at.%) | [In] (at.%) | GGI | [S] (at.%) | [Se] (at.%) | SSSe |
|---|---|---|---|---|---|---|---|
| **Standard CISSe cell** | APT-1 | 0.058 | 27.6 | 0.0021 | 18.9 | 30.6 | 0.38 |
| | APT-2 | 0.067 | 27.6 | 0.0024 | 15.5 | 33.2 | 0.31 |
| | APT-3 | 0.079 | 28.5 | 0.0027 | 17.5 | 31 | 0.36 |
| | Average | **0.068** | 27.9 | 0.0024 | **17.3** | 31.6 | 0.35 |
| **High CISSe cell** | APT-1 | 0.125 | 27.8 | 0.0044 | 26.1 | 22 | 0.54 |
| | APT-2 | 0.22 | 27.4 | 0.0079 | 28.7 | 20.4 | 0.58 |
| | APT-3 | 0.23 | 27.8 | 0.0082 | 28.5 | 19.7 | 0.59 |
| | Average | **0.19** | 27.6 | 0.0068 | **27.8** | 20.7 | 0.57 |
| **High CIGSSe cell** | APT-1 | 1.3 | 27.2 | 0.0456 | 31.6 | 17.8 | 0.63 |
| | APT-2 | 1.93 | 26.7 | 0.0674 | 31.8 | 17.7 | 0.64 |
| | APT-3 | 1.89 | 25 | 0.0702 | 39.5 | 9.6 | 0.80 |
| | Average | **1.70** | 26.3 | 0.0611 | **34.3** | 15.03 | 0.69 |

Surprisingly, no changes in the In and Cu compositions were detected at the absorber surface for all three cells. This is contrary with the typical Cu-depleted CIGSe surface observed previously by APT for various types of buffers such as CdS[40-41], In$_2$S$_3$[44, 49], and mixed ZnS-In$_2$S$_3$[50]. This suggests, hence, that the main role at the CIGSe surface is played by S and not by Cu.

### 3.4 Study of the band alignment through DFT calculation

To fully understand the effect of S on the cell performance, one needs to understand also the impact of S on the electronic band structure of the cell, especially in the region of the *p-n* junction. Since our TEM experiments showed that the In$_x$S$_y$:Na buffer contains tetragonal In$_2$S$_3$ and/or NaIn$_3$S$_5$, we considered the NaIn$_3$S$_5$ phase for the DFT calculations because this phase contains inherently about 11 at.% of Na, which is close to the Na content determined by APT for the standard CISSe sample. Hence, band-offset calculations based on DFT were performed to highlight the change in band alignment between the buffer NaIn$_3$S$_5$ and the absorber when S and Ga were added subsequently to the CuInSe$_2$ or Cu(In,Ga)Se$_2$ absorber materials.

Figure 8 shows the theoretically calculated band alignment of NaIn$_3$S$_5$ with CIGSe-based chalcopyrite compounds. In chalcopyrite systems, Cu *d* orbitals are close to the VBM. Therefore, there is a large *p-d* repulsion between Cu *d* and anion *p* orbitals. S *p* orbitals (*2p*) are deeper in energy compared to Se *p* orbitals (*3p*). Hence, the energy difference between Cu *d* and S *p* states is smaller than that between Cu *d* and Se *p*, suggesting stronger *p-d* interaction in S-containing chalcopyrite. Therefore, in contrast to II-VI systems[12], it is expected that by substituting Se atoms with S in CuInSe$_2$ the position of VBM does not decrease considerably. As shown in Figure 7, by substituting Se anion sites with S, the downward shift in the VBM of CuIn(S$_{0.25}$,Se$_{0.75}$)$_2$, CuIn(S,Se)$_2$, and CuIn(S$_{0.75}$,S$_{0.25}$)$_2$ with respect to CuInSe$_2$ are only 0.02, 0.04, and 0.08 eV, respectively. However, the conduction bands of CuIn(S$_{0.25}$,Se$_{0.75}$)$_2$, CuIn(S$_{0.5}$,Se$_{0.5}$)$_2$, and CuIn(S$_{0.75}$,S$_{0.25}$)$_2$ mixed-anion systems are 0.11, 0.18, and 0.25 eV higher than the CBM of CuInSe$_2$, respectively.

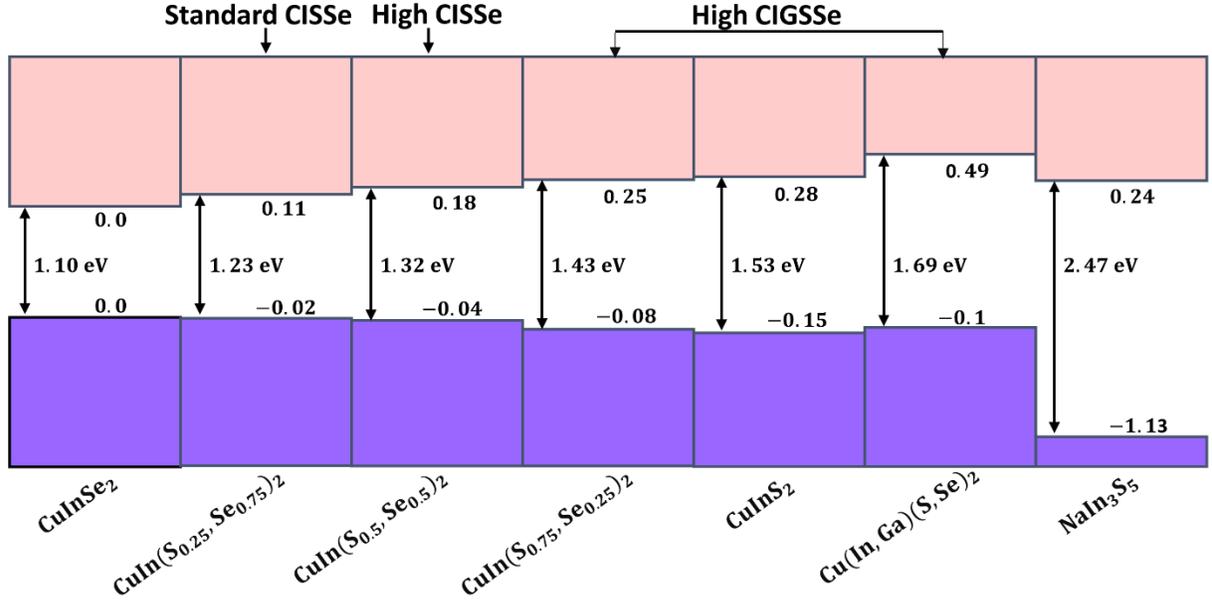

*Figure 8. Calculated band line-ups between CuInSe$_2$, CuIn(S$_{0.25}$,Se$_{0.75}$)$_2$, CuIn(S$_{0.5}$,Se$_{0.5}$)$_2$, CuIn(S$_{0.75}$,S$_{0.25}$)$_2$, CuInS$_2$, Cu(In,Ga)(S,Se)$_2$, and NaIn$_3$S$_5$. The lower and upper numbers indicate positions of VBM and CBM of each compound with respect to the VBM and CBM of CuInSe$_2$, given in eV. Black arrows indicate the band gap of each material.*

It is interesting to note that when we alloy CuInSe$_2$ with both S (12.5 at.%) and Ga (12.5 at.%) atoms, the valence band offset remains small (0.10 eV) and most of the band offset occurs in the conduction band (0.49 eV). The electronic origin of this feature can be rationalized by considering the impact of *p-d* hybridization on valence band levels and cation size on conduction band levels. As reported by Ghorbani et al.[51], the band offset between CuInSe$_2$ and CuGaSe$_2$ follows the common anion rule alignment, demonstrating a negligible valence band offset of 0.13 eV and a much larger conduction band offset of 0.71 eV. This feature can be traced back to the fact that substitution of In atoms by smaller atoms such as Ga reduces the cell volume, which increases the band gap and moves the conduction band states upward. Yet, one need to mention that the Ga content at the CISSe surface measured by APT for all three investigated cells (Table 2) is well below 12.5 at.%, value used in the DFT calculations.

### 3.5 Impact of S and Ga on the performance of buffer/absorber heterointerface

When considering the interface region of a heterojunction solar cell in view of possible interface recombination, it is instructive to look at the quantity $E_{p,z=0}$ which is the energetic distance of the Fermi energy to the valence band maximum at the absorber/emitter interface (see Figure 9 for illustration). For a solar cell with an inverted absorber surface ($E_{p,z=0} > E_{n,z=0}$), interface recombination increases with smaller $E_{p,z=0}$ according to sections 3.2 and 3.3 of Ref. [28]. The quantity $E_{p,z=0}$ depends both on the *conduction band offset* between the buffer and absorber (denoted as $\Delta E_c$; see section 3.2 in Ref. [28]) and the doping ratio between absorber and emitter (buffer/window; section 3.3). When doping in the emitter is kept constant, $E_{p,z=0}$ depends on the *hole "p" density* in the absorber. The value of $E_{p,z=0}$ decreases with smaller *conduction band offset $\Delta E_c$* and increases with lower *hole "p" density* in the absorber. Both effects need to be balanced.

Our results indicate that $\Delta E_c(CuIn(S_{0.25},Se_{0.75})_2/Na\,In_3S_5)$ calculated for the standard CISSe cell is +0.13 eV (see Figure 8), forming a modest 'spike-like' conduction band offset in agreement with the work of Insignares-Cuello et al.[6] and Hauschild et al. [52], who observed a rather flat band transition between the absorber and InS:Na buffer. Adding S and Ga to the $CuInSe_2$ compound will shift the conduction band minimum to higher energies resulting in smaller band offsets; i.e. only +0.06 eV for the high CISSe cell and even ~ -0.07 eV for the high CIGSSe cell (see the offsets in Figure 9). As stated above, the smaller band offset leads to smaller $E_{p,z=0}$. The second parameter is the *hole "p" density* at the interface region. From the C-V and APT experiments it is indicated that it can be the high S content in high CISSe and high CIGSSe that reduces the hole density (or *p*-doping) in the absorber. This argument is based on the work of Walter et al.[53] who find lower p-type doping with increasing Sulfur content. Now, lower p-doping means larger $E_{p,z=0}$ as then the absorber exhibits more band bending. This explains why in Figure 9(b) the $E_{p,z=0}$ becomes larger when adding S.

Based on $E_{p,z=0}$ which depends both on *$\Delta E_c$* and the *hole "p" density* in the interface region of the absorber, the investigated cells can be described as follows. The Standard CISSe cell is characterized by a relative high $\Delta E_c$ value with a spike-like behavior as proven by DFT. From the APT experiments we learned that the S content at the absorber surface is rather low (i.e. high p-doping at the absorber surface) which explains the rather low $E_{p,z=0}$ value as given in the simulated band diagram in Figure 9a. With this low $E_{p,z=0}$ the recombination rate at the interface is high for the standard CISSe sample and may even dominate the device recombination. Interface recombination under red light illumination is validated experimentally by the falling $V_{OC}(t)$ transient. The combination of a spike, a large $E_{n,z=0}$ (due to small $E_{p,z=0}$) and (assumed) interface defect states also allows to explain the kink in the red light JV curve: Photogenerated electrons cannot pass the $\Delta E_c$ barrier and recombine at the interface. The photocurrent is reduced forming the kink. Under white light illumination, however, photodoping of the buffer changes the doping ratio, decreases $E_{n,z=0}$, allows transport of electrons over the $\Delta E_c$ barrier, and reduces interface recombination. We presume that interface recombination to some extend still adds to the total recombination current even under white light illumination where it is found in the transients that $dV_{oc}(t)/dt \approx 0$. Now, the situation is different for the high S content CI(G)SSe cells where the high S content as given by APT reduces the p-doping at the absorber surface. As explained above, this leads to a larger $E_{p,z=0}$ value as schematized in Figure 9b,c, i.e. to fewer holes available at the interface. Now, interface recombination is suppressed. This is in agreement with the rising $V_{OC}(t)$ transient in Figure 4. A rising $V_{oc}(t)$ transient indicates *bulk recombination* to be the main factor that limits the device performance with the two possibilities: recombination in the space-charge-region and/or in the quasi-neutral region. Interface recombination may be further reduced in the high S samples, since, as revealed by *ab-inito* calculations (see Figure 8), the resultant bandgap increase is partly realized by a reduced valence band maximum, further reducing the hole density at the interface in agreement with the work of Kim et al.[54]. Since the S content at the absorber surface in high CIGSSe (34.3 at.%; very low p-doping) is even higher than that in high CISSe (27.7 at.%; low p-doping), the $E_{p,z=0}$ value for high CIGSSe (Figure 9c) is qualitatively higher than the one for high CISSe. Thus, the hole density at the interface for high CIGSSe is much more reduced indicating that the probability for interface recombination to take place is even lower in the high CIGSSe sample. Moreover, this scenario might correspond to a *type-inverted interface* since $E_{p,z=0} > E_{n,z=0}$. This fits well with the formation of the Se-doped CuInS$_2$ phase at the absorber surface as proven by APT and TEM, since the CuInS$_2$ phase is supposed to have less p-type character due to the presence of a high concentration of S vacancies $V_S^{2-}$ acting as compensating donors[55].

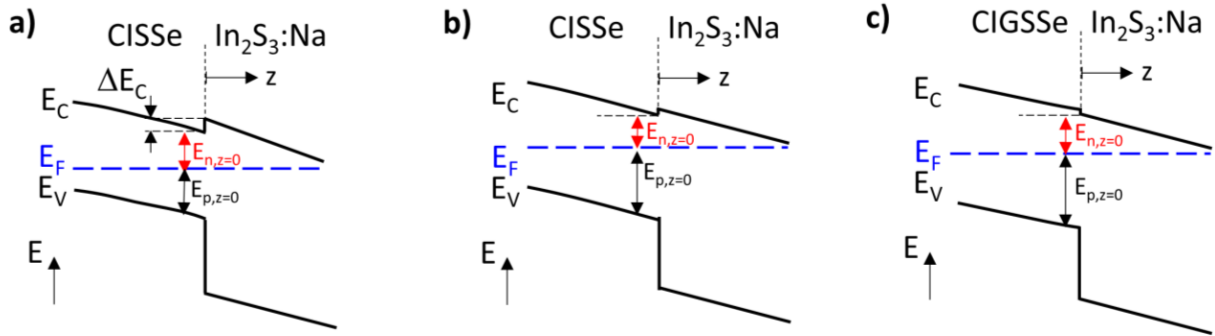

*Figure 9. Schematic drawings of the energy-band diagrams for the cells based on all experimental and theoretical results.* *The studied cells are a) standard CISSe, b) high CISSe and c) high CIGSSe. The standard CISSe is limited by interface recombination under red light due to the almost equal hole p and electron n density, i.e. by almost equal $E_{p,z=0}$ and $E_{n,z=0}$ values. The band diagram may change towards smaller $E_{n,z=0}$ upon white light illumination. On contrary, the hole density p at the absorber surface (high $E_{p,z=0}$ value) is reduced due to the higher S content at the absorber surface for high CISSe and high CIGSSe cells. This explains why the interface recombination for these cells is reduced and, hence, bulk recombination becomes the dominant recombination mechanism even under red light illumination. By further increasing the S content at the absorber surface for high CIGSSe sample, the p-doping is strongly reduced so that the $E_{p,z=0} > E_{n,z=0}$. This corresponds to a scenario where the interface is type-inverted. The band diagrams were simulated using the program AforsHet.*

Finally, the effect of Ga at the absorber surface needs to be discussed. Ga in CIGSSe is known to have three main effects: it increases the p-doping, reduces the tetragonal distortion (i.e. δ=c-2a where a and c are the lattice parameters) [56] and increases the band gap. An increased p-doping cannot be observed in the C-V data of the high CIGSSe sample, probably because the Ga effect is smaller than the S effect. Second, due to the low Ga concentration in the sampled region of the absorber, the effect of tetragonal distortion reduction on the defect density and thus on the bulk recombination rate may be small. Hence, the bandgap effect remains to explain the increase in $V_{oc}$ of the high CIGSSe sample. The Shockley-Queisser theory predicts that the limiting efficiency of ideal solar cells increases for a higher bandgap up to the value of around 1.2 eV. However, for solar cells limited by recombination in the SCR this effect is even larger (see Figure 3.1 in Ref. [28]): The optimum band gap (depending on the charge carrier lifetime) may be shifted to 1.4 eV. Thus, the (slightly) increased $V_{oc}$ of the high CIGSSe sample here may be interpreted as due to the (slightly) larger bandgap of the absorber in the front region of the layer. The increased bandgap would also explain the (slightly) reduced $J_{sc}$ of this device.

## 4. Conclusions

In the present work we studied the joint effect of Ga and S grading on the buffer/absorber heterojunction properties using both, experimental and theoretical investigations. We found superior electrical properties of the buffer/absorber heterojunction when the intense sulphurization of the absorber surface is accompanied by a very modest Ga-grading. More precisely, we obtained a very strong reduction of the p-doping by C-V experiments together with a little negative band offset ($\Delta E_C$=-0.07 eV from DFT calculations). These results fit well with the $V_{OC}(t)$ transient measurements where a weak sulphurization leads to interface recombination as dominant recombination process under red light illumination, whereas the intense sulphurization leads to bulk recombination. Moreover, a S-rich CuIn(Se,S)$_2$ type-inverted layer was detected at the absorber surface. This particular configuration makes the surface to repel holes and attract electrons explaining the reduced non-radiative recombination and the 10 % increase in $V_{OC}$. Hence, the present work opens new avenues for achieving outstanding heterojunction properties and, hence, cell performance when using intense sulphurization.

## Acknowledgements

The authors thank Thomas Niesen and Patrick Eraerds from Avancis GmbH for the samples provided and for the GDOES, I-V, and CV experiments. The present work was conducted within the EFFCIS consortium and financed by Federal Ministry for Economic Affairs and Energy-BMWi (0324076A-G).